%%%%%%%%%%%%%%%%%%%%%%%%%%%%%%%%%%%%%%%%%%%%%%%%%%%%%%%%%%%%%%%%%%%%
%  TeX Definitions                                                 %
%%%%%%%%%%%%%%%%%%%%%%%%%%%%%%%%%%%%%%%%%%%%%%%%%%%%%%%%%%%%%%%%%%%%

\newif\iffigs\figstrue
% Uncomment the next line if you do not want the figures
%\figsfalse

%
% the following is to use blackboard bold fonts --
\let\useblackboard=\iftrue
%
% activate this if you don't have them.
%\let\useblackboard=\iffalse
%
% You might also need to remove this line.
\newfam\black

\input harvmac.tex

\iffigs
  \input epsf
\else
  \message{No figures will be included.  See TeX file for more
information.}
\fi

\def\Title#1#2{\rightline{#1}
\ifx\answ\bigans\nopagenumbers\pageno0\vskip1in%
\baselineskip 15pt plus 1pt minus 1pt
\else%\special{papersize=11in,8.5in}%
\def\listrefs{\footatend\vskip 1in\immediate\closeout\rfile\writestoppt
\baselineskip=14pt\centerline{{\bf References}}\bigskip{\frenchspacing%
\parindent=20pt\escapechar=` \input
refs.tmp\vfill\eject}\nonfrenchspacing}
\pageno1\vskip.8in\fi \centerline{\titlefont #2}\vskip .5in}

\ifx\answ\bigans\def\tcbreak#1{}\else\def\tcbreak#1{\cr&{#1}}\fi
\useblackboard
\message{If you do not have msbm (blackboard bold) fonts,}
\message{change the option at the top of the tex file.}
\font\blackboard=msbm10 %scaled \magstep1
\font\blackboards=msbm7
\font\blackboardss=msbm5
%\newfam\black
\textfont\black=\blackboard
\scriptfont\black=\blackboards
\scriptscriptfont\black=\blackboardss
\def\Bbb#1{{\fam\black\relax#1}}
\else
\def\Bbb#1{{\bf #1}}
\fi
% *************************************
%
\def\yboxit#1#2{\vbox{\hrule height #1 \hbox{\vrule width #1
\vbox{#2}\vrule width #1 }\hrule height #1 }}
\def\fillbox#1{\hbox to #1{\vbox to #1{\vfil}\hfil}}
\def\ybox{{\lower 1.3pt \yboxit{0.4pt}{\fillbox{8pt}}\hskip-0.2pt}}
\def\np#1#2#3{Nucl. Phys. {\bf B#1} (#2) #3}

\def\comments#1{}

\def\half{{1\over 2}}

\def\ket#1{|#1\rangle}

\def\a{\alpha}

\def\II{\relax{I\kern-.07em I}}

\def\IZ{\relax\ifmmode\mathchoice
{\hbox{\cmss Z\kern-.4em Z}}{\hbox{\cmss Z\kern-.4em Z}}
{\lower.9pt\hbox{\cmsss Z\kern-.4em Z}}
{\lower1.2pt\hbox{\cmsss Z\kern-.4em Z}}\else{\cmss Z\kern-.4em
Z}\fi}
\def\IB{\relax{\rm I\kern-.18em B}}

\def\ID{\relax{\rm I\kern-.18em D}}
\def\IE{\relax{\rm I\kern-.18em E}}
\def\IF{\relax{\rm I\kern-.18em F}}
\def\IG{\relax\hbox{$\inbar\kern-.3em{\rm G}$}}
\def\IGa{\relax\hbox{${\rm I}\kern-.18em\Gamma$}}
\def\IH{\relax{\rm I\kern-.18em H}}
\def\II{\relax{\rm I\kern-.18em I}}
\def\IK{\relax{\rm I\kern-.18em K}}
\def\IP{\relax{\rm I\kern-.18em P}}
%\def\IX{\relax{\rm X\kern-.01em X}}
%this doesn't work

\useblackboard
\def\IZ{\relax\Bbb{Z}}
\fi

\font\cmss=cmss10 \font\cmsss=cmss10 at 7pt
\def\IR{\relax{\rm I\kern-.18em R}}

\def\BR{\IR}
\def\BZ{\IZ}
\def\BR{\IR}

%%%%%%%%%%%%%%%%%%%%%%%%%%%%%%%%%%%%%%%%%%%%%%%%%%%%%%%%%%%%%%%%%%%%%%%%%%%%
%                    F I G U R E S                                         %
%%%%%%%%%%%%%%%%%%%%%%%%%%%%%%%%%%%%%%%%%%%%%%%%%%%%%%%%%%%%%%%%%%%%%%%%%%%%
%\figsfalse

\iffigs
  \input epsf
\else
  \message{No figures will be included.  See TeX file for more
information.}
\fi

%%% \iffigs
%%% \midinsert
%%% $$\vbox{\centerline{\epsfxsize=4in\epsfbox{figa.eps}}
%%% \centerline{Figure 1. $E_2$ and related theories.}}$$
%%% \endinsert
%%% \fi

%%%%%%%%%%%%%%%%%%%%%%%%%%%%%%%%%%%%%%%%%%%%%%%%%%%%%%%%%%%%%%%%%%%%%%%%%%%%
%  Tables - taken from hep-th/9605200                                      %
%%%%%%%%%%%%%%%%%%%%%%%%%%%%%%%%%%%%%%%%%%%%%%%%%%%%%%%%%%%%%%%%%%%%%%%%%%%%

%
% S-Tables Macro
%
\message{S-Tables Macro v1.0, ACS, TAMU (RANHELP@VENUS.TAMU.EDU)}
%
% Help Text
%
\newhelp\stablestylehelp{You must choose a style between 0 and 3.}%
\newhelp\stablelinehelp{You 
should not use special hrules when stretching
a table.}%
\newhelp\stablesmultiplehelp{You have tried to place an S-Table  
inside another
S-Table.  I would recommend not going on.}%
%
% Line Thicknesses (Values)
%
\newdimen\stablesthinline
\stablesthinline=0.4pt
\newdimen\stablesthickline
\stablesthickline=1pt
%
% Border and Internal Line Thicknesses
%
\newif\ifstablesborderthin
\stablesborderthinfalse
\newif\ifstablesinternalthin
\stablesinternalthintrue
\newif\ifstablesomit
\newif\ifstablemode
\newif\ifstablesright
\stablesrightfalse
%
% Save Registers
%
\newdimen\stablesbaselineskip
\newdimen\stableslineskip
\newdimen\stableslineskiplimit
%
% Counters
%
\newcount\stablesmode
\newcount\stableslines
\newcount\stablestemp
\stablestemp=3
\newcount\stablescount
\stablescount=0
\newcount\stableslinet
\stableslinet=0
%
% Table Style Selection
%
% 0 - Centered
% 1 - Left Justified
% 2 - Right Justified
% 3 - Not Justified
%
\newcount\stablestyle
\stablestyle=0
%
% Element Buffering Definitions
%
\def\stablesleft{\quad\hfil}%
\def\stablesright{\hfil\quad}%
%
% Vertical Bar Activation
%
\catcode`\|=\active%
%
% Strut Control
%
\newcount\stablestrutsize
\newbox\stablestrutbox
\setbox\stablestrutbox=\hbox{\vrule height10pt depth5pt width0pt}
\def\stablestrut{\relax\ifmmode%
                         \copy\stablestrutbox%
                       \else%
                         \unhcopy\stablestrutbox%
                       \fi}%
%
% Misc. Internal Stuff
%
\newdimen\stablesborderwidth
\newdimen\stablesinternalwidth
\newdimen\stablesdummy
\newcount\stablesdummyc
\newif\ifstablesin
\stablesinfalse
%
% Table Macros
%
\def\begintable{\stablestart%
  \stablemodetrue%
  \stablesadj%
  \halign%
  \stablesdef}%
\def\stablesadj{%
  \ifcase\stablestyle%
    \hbox to \hsize\bgroup\hss\vbox\bgroup%
  \or%
    \hbox to \hsize\bgroup\vbox\bgroup%
  \or%
    \hbox to \hsize\bgroup\hss\vbox\bgroup%
  \or%
    \hbox\bgroup\vbox\bgroup%
  \else%
    \errhelp=\stablestylehelp%
    \errmessage{Invalid style selected, using default}%
    \hbox to \hsize\bgroup\hss\vbox\bgroup%
  \fi}%
\def\stablesend{\egroup%
  \ifcase\stablestyle%
    \hss\egroup%
  \or%
    \hss\egroup%
  \or%
    \egroup%
  \or%
    \egroup%
  \else%
    \hss\egroup%
  \fi}%
\def\stablestart{%
  \ifstablesin%
    \errhelp=\stablesmultiplehelp%
    \errmessage{An S-Table cannot be placed within an S-Table!}%
  \fi
  \global\stablesintrue%
  \global\advance\stablescount by 1%
  \message{<S-Tables Generating Table \number\stablescount}%
  \begingroup%
  \stablestrutsize=\ht\stablestrutbox%
  \advance\stablestrutsize by \dp\stablestrutbox%
  \ifstablesborderthin%
    \stablesborderwidth=\stablesthinline%
  \else%
    \stablesborderwidth=\stablesthickline%
  \fi%
  \ifstablesinternalthin%
    \stablesinternalwidth=\stablesthinline%
  \else%
    \stablesinternalwidth=\stablesthickline%
  \fi%
  \tabskip=0pt%
  \stablesbaselineskip=\baselineskip%
  \stableslineskip=\lineskip%
  \stableslineskiplimit=\lineskiplimit%
  \offinterlineskip%
  \def\borderrule{\vrule width \stablesborderwidth}%
  \def\internalrule{\vrule width \stablesinternalwidth}%
  \def\thinline{\noalign{\hrule height \stablesthinline}}%
  \def\thickline{\noalign{\hrule height \stablesthickline}}%
  \def\trule{\omit\leaders\hrule height \stablesthinline\hfill}%
  \def\ttrule{\omit\leaders\hrule height \stablesthickline\hfill}%
  \def\tttrule##1{\omit\leaders\hrule height ##1\hfill}%
  \def\stablesel{&\omit\global\stablesmode=0%
    \global\advance\stableslines by 1\borderrule\hfil\cr}%
  \def\el{\stablesel&}%
  \def\elt{\stablesel\thinline&}%
  \def\eltt{\stablesel\thickline&}%
  \def\elttt##1{\stablesel\noalign{\hrule height ##1}&}%
  \def\elspec{&\omit\hfil\borderrule\cr\omit\borderrule&%
              \ifstablemode%
              \else%
                \errhelp=\stablelinehelp%
                \errmessage{Special ruling will not display properly}%
              \fi}%
  \def\stmultispan##1{\mscount=##1 \loop\ifnum\mscount>3
\stspan\repeat}%
  \def\stspan{\span\omit \advance\mscount by -1}%
  \def\multicolumn##1{\omit\multiply\stablestemp by ##1%
     \stmultispan{\stablestemp}%
     \advance\stablesmode by ##1%
     \advance\stablesmode by -1%
     \stablestemp=3}%
  \def\multirow##1{\stablesdummyc=##1\parindent=0pt\setbox0\hbox\bgroup%
    \aftergroup\emultirow\let\temp=}
  \def\emultirow{\setbox1\vbox to\stablesdummyc\stablestrutsize%
    {\hsize\wd0\vfil\box0\vfil}%
    \ht1=\ht\stablestrutbox%
    \dp1=\dp\stablestrutbox%
    \box1}%
  
\def\stpar##1{\vtop\bgroup\hsize ##1%
     \baselineskip=\stablesbaselineskip%
     \lineskip=\stableslineskip%
      
\lineskiplimit=\stableslineskiplimit\bgroup\aftergroup\estpar\let\temp=}%
  \def\estpar{\vskip 6pt\egroup}%
  \def\stparrow##1##2{\stablesdummy=##2%
     \setbox0=\vtop to ##1\stablestrutsize\bgroup%
     \hsize\stablesdummy%
     \baselineskip=\stablesbaselineskip%
     \lineskip=\stableslineskip%
     \lineskiplimit=\stableslineskiplimit%
     \bgroup\vfil\aftergroup\estparrow%
     \let\temp=}%
  \def\estparrow{\vfil\egroup%
     \ht0=\ht\stablestrutbox%
     \dp0=\dp\stablestrutbox%
     \wd0=\stablesdummy%
     \box0}%
  \def|{\global\advance\stablesmode by 1&&&}%
  \def\|{\global\advance\stablesmode by 1&\omit\vrule width 0pt%
         \hfil&&}%
  \def\vt{\global\advance\stablesmode by 1&\omit\vrule width  
\stablesthinline%
          \hfil&&}%
  \def\vtt{\global\advance\stablesmode by 1&\omit\vrule width
\stablesthickline%
          \hfil&&}%
  \def\vttt##1{\global\advance\stablesmode by 1&\omit\vrule width ##1%
          \hfil&&}%
  \def\vtr{\global\advance\stablesmode by 1&\omit\hfil\vrule width%
           \stablesthinline&&}%
  \def\vttr{\global\advance\stablesmode by 1&\omit\hfil\vrule width%
            \stablesthickline&&}%
  \def\vtttr##1{\global\advance\stablesmode by 1&\omit\hfil\vrule  
width ##1&&}%
  \stableslines=0%
  \stablesomitfalse}
\def\stablesdef{\bgroup\stablestrut\borderrule##\tabskip=0pt plus 1fil%
  &\stablesleft##\stablesright%
  &##\ifstablesright\hfill\fi\internalrule\ifstablesright\else\hfill\fi%
  \tabskip 0pt&&##\hfil\tabskip=0pt plus 1fil%
  &\stablesleft##\stablesright%
  &##\ifstablesright\hfill\fi\internalrule\ifstablesright\else\hfill\fi%
  \tabskip=0pt\cr%
  \ifstablesborderthin%
    \thinline%
  \else%
    \thickline%
  \fi&%
}%
\def\endtable{\advance\stableslines by 1\advance\stablesmode by 1%
   \message{- Rows: \number\stableslines, Columns:   
\number\stablesmode>}%
   \stablesel%
   \ifstablesborderthin%
     \thinline%
   \else%
     \thickline%
   \fi%
   \egroup\stablesend%
\endgroup%
\let |=\vert % Added by me, to restore definition of |
\global\stablesinfalse}
%
% end of STABLES.TEX

%\input stables.tex
%\input tab.tex

%%%%%%%%%%%%%%%%%%%%%%%%%%%%%%%%%%%%%%%%%%%%%%%%%%%%%%%%%%%%%%%%%%%%%%%%%%%%
%                    Definitions from LaTeX                                %
%%%%%%%%%%%%%%%%%%%%%%%%%%%%%%%%%%%%%%%%%%%%%%%%%%%%%%%%%%%%%%%%%%%%%%%%%%%%

%%%
%%% All those have problems with Font \rm
%%%

\def\lim{{lim}}

%%%%%%%%%%%%%%%%%%%%%%%%%%%%%%%%%%%%%%%%%%%%%%%%%%%%%%%%%%%%%%%%%%%%%%%%%%%%
%                    My definitions                                        %
%%%%%%%%%%%%%%%%%%%%%%%%%%%%%%%%%%%%%%%%%%%%%%%%%%%%%%%%%%%%%%%%%%%%%%%%%%%%
\input epsf

\def\SUSY#1{{{\cal N}= {#1}}}                   % N=? SUSY
\def\lbr{{\lbrack}}                             % [
\def\rbr{{\rbrack}}                             % ]

\def\wdg{{\wedge}}                              % wedge product

                              % inverse
                           % O(x)

\def\MR#1{{{\BR}^{#1}}}               % Real numbers
               % Complex numbers

%%% \def\MR#1{{{\bf R}^{#1}}}               % Real numbers
%%% \def\MC#1{{{\bf C}^{#1}}}               % Complex numbers
\def\MR#1{{{\BR}^{#1}}}               % Real numbers
               % Complex numbers
\def\MS#1{{{\bf S}^{#1}}}               % Circle, sphere,...
               % disk, ball,...
\def\MT#1{{{\bf T}^{#1}}}               % Torus
              % CP
               % Ruled surface F_n

             % Patch
                    % line-bundle
\def\px#1{{\partial_{#1}}}              % derivative

                 % Left large bracket
                % Right large bracket
              % SL(*,Z)

                             % identity matrix

      % commutator
               % anti-commutator

           % expectation value
    % expectation value of trace

      % trace
\def\trp#1{{{\rm tr}\{ {#1} \} }}            % trace
            % Trace
\def\trr#1#2{{{\rm tr}_{#1}\{ {#2} \} }}            % trace in a rep
            % Trace in a rep

\def\rep#1{{{\bf {#1}}}}                      % representation
                  % Imaginary
                  % Imaginary

                  % Imaginary
                  % Imaginary

%\def\widebar#1{{\bar{#1}}}                    % Wide bar
                    % Wide bar
                 % Pauli matrix

\def\Ol#1{{ {\cal O}({#1}) }}                      % correction O()
\def\Hil{{\cal H}}                                 % Hilbert space
\def\MSO{{\cal M}}                                 % Parameter space

                      % Hodge star
                         % sign
\def\hepth#1{{\it hep-th/{#1}}}

%%%%%%%%%%%%%%%%%%%%%%%%%%%%%%%%%%%%%%%%%%%%%%%%%%%%%%%%%%%%%%%%%%%%%%%%%%%%
%                    Greek                                                 %
%%%%%%%%%%%%%%%%%%%%%%%%%%%%%%%%%%%%%%%%%%%%%%%%%%%%%%%%%%%%%%%%%%%%%%%%%%%%
\def\u{{\mu}}
\def\v{{\nu}}
\def\b{{\beta}}

\def\da{{\dot{\a}}}
\def\db{{\dot{\b}}}

\def\lam{{\lambda}}

%%% \def\Dsh{{D\!\!\!\slash}}     % D slash

%%%%%%%%%%%%%%%%%%%%%%%%%%%%%%%%%%%%%%%%%%%%%%%%%%%%%%%%%%%%%%%%%%%%%%%%%%%%
%     Special Purpose  Definitions                                         %
%%%%%%%%%%%%%%%%%%%%%%%%%%%%%%%%%%%%%%%%%%%%%%%%%%%%%%%%%%%%%%%%%%%%%%%%%%%%
                            % 2\times 2  J

%%% \def\ww{{\widebar{w}}}
%%% \def\wu{{\widebar{u}}}
%%% \def\wv{{\widebar{v}}}

          % type-IIB string coupling

%%% \def\vpp{{\vec{p}{}'}}
%%% \def\vqp{{\vec{q}{}'}}
%%% \def\vwp{{\vec{w}{}'}}

       % String oscillator
      % fermionic oscillator
\def\MHT#1{{\widehat{{\bf T}}^{#1}}}               % Torus

      % blow-up
\def\wH{{\widetilde{H}}}      % another Hamiltonian

%%%%%%%%%%%%%%%%%%%%%%%%%%%%%%%%%%%%%%%%%%%%%%%%%%%%%%%%%%%%%%%%%%%%%%%%%%%%
%     Draftmode                                                            %
%%%%%%%%%%%%%%%%%%%%%%%%%%%%%%%%%%%%%%%%%%%%%%%%%%%%%%%%%%%%%%%%%%%%%%%%%%%%
%%% \draftmode
%%%%%%%%%%%%%%%%%%%%%%%%%%%%%%%%%%%%%%%%%%%%%%%%%%%%%%%%%%%%%%%%%%%%%%%%%%%%
%                    TITLE PAGE                                            %
%%%%%%%%%%%%%%%%%%%%%%%%%%%%%%%%%%%%%%%%%%%%%%%%%%%%%%%%%%%%%%%%%%%%%%%%%%%%

%
\Title{ \vbox{\baselineskip12pt\hbox{hep-th/9709139, PUPT-1715}}}
{\vbox{
\centerline{On The M(atrix)-Model For M-Theory On $T^6$}}}
\centerline{Ori J. Ganor}
\smallskip
\smallskip
\centerline{Department of Physics, Jadwin Hall}
\centerline{Princeton University}
\centerline{Princeton, NJ 08544, USA}
\centerline{\tt origa@puhep1.princeton.edu}

%%% \bigskip
%%% \bigskip
%%% \centerline{and}
%%% \bigskip
%%% \bigskip
%%% \centerline{FILL HERE}
%%% \smallskip
%%% \smallskip

\bigskip
\bigskip
\noindent
We study consistency conditions on a M(atrix)-model
which would describe M-theory on $T^6$.
We argue that there is
a limit in moduli space for which it becomes a 6+1D theory
and study the low-energy description of extended objects in 
the decompactified limit.
We discuss the requirements from a M(atrix)-model which would
describe such an $E_{6(6)}$ theory.
We suggest that it could be a 5+1D theory and that a 1+1D
theory with $(0,4)$ supersymmetry might be the M(atrix)-model
for the M(atrix)-model of the $E_{6(6)}$ theory.

\Date{September, 1997}

%%%%%%%%%%%%%%%%%%%%%%%%%%%%%%%%%%%%%%%%%%%%%%%%%%%%%%%%%%%%%%%%%%%%
%  B I B L I O G R A P H Y                                         %
%%%%%%%%%%%%%%%%%%%%%%%%%%%%%%%%%%%%%%%%%%%%%%%%%%%%%%%%%%%%%%%%%%%%

\lref\SeiVBR{N. Seiberg,
  {\it ``Matrix Description of M-theory on $T^5$ and $T^5/Z_2$,''}
  \hepth{9705221}}

\lref\BFSS{T. Banks, W. Fischler, S.H. Shenker and L. Susskind,
  {\it ``M Theory As A Matrix Model: A Conjecture,''} \hepth{9610043}}

\lref\DVVQ{R. Dijkgraaf, E. Verlinde, H. Verlinde,
  {\it ``BPS Quantization Of The 5-Brane,''} 
  \np{486}{97}{77}, \hepth{9604055}}
\lref\DVVS{R. Dijkgraaf, E. Verlinde, H. Verlinde,
  {\it ``BPS Spectrum Of The 5-Brane And Black-Hole Entropy,''} 
  \np{486}{97}{89}, \hepth{9604055}}
%%% \lref\DVVX{R. Dijkgraaf, E. Verlinde, H. Verlinde,
%%%   On the deformation of the sigma model of instantons.}

\lref\WitCOM{ E. Witten,
  {\it ``Some Comments on String Dynamics,''}
  contributed to Strings '95, \hepth{9507121}}

\lref\StrOPN{A. Strominger,
  {\it ``Open p-Branes,''} \hepth{9512059}}

\lref\WitQHB{E. Witten,
  {\it ``On The Conformal Field Theory Of The Higgs Branch,''}
  \hepth{9707093}}

\lref\SusANO{L. Susskind,
  {\it ``Another Conjecture about M(atrix) Theory,''} \hepth{9704080}}

\lref\BanksR{T. Banks,
  {\it ``The State of Matrix Theory,''} \hepth{9706168}}

\lref\SWSIXD{N. Seiberg and E. Witten,
  {\it ``Comments On String Dynamics In Six-Dimensions,''}
  \hepth{9603003}}

\lref\Wati{W. Taylor,
  {\it ``D-brane field theory on compact spaces,''} \hepth{9611042}}

\lref\EGKR{S. Elitzur, A. Giveon, D. Kutasov and E. Rabinovici,
  {\it ``Algebraic Aspects of Matrix Theory on $T^d$,''}
  \hepth{9707217}}

\lref\SeiSet{N. Seiberg and S. Sethi,
  {\it ``Comments on Neveu-Schwarz Five-Branes,''} \hepth{9708085}}

\lref\FHRS{W. Fischler, E. Halyo, A. Rajaraman, and L. Susskind,
  {\it ``The Incredible Shrinking Torus,''} \hepth{9703102}}

\lref\WitVAR{E. Witten,
  {\it ``String Theory Dynamics in Various Dimensions,''}
  \np{443}{95}{85}, \hepth{9503124}}

\lref\SeiNTS{N. Seiberg,
  {\it ``Notes on Theories with 16 Supercharges,''} \hepth{9705117}}

\lref\LMS{A. Losev, G. Moore and S.L. Shatashvili,
  {\it ``M\&m's,''} \hepth{9707250}}

\lref\HL{A. Hanany and G. Lifschytz,
  {\it ``M(atrix) Theory on $T^6$ and a m(atrix) Theory Description of KK
  Monopoles,''} \hepth{9708037}}

\lref\HulTow{C. M. Hull and P.K. Townsend,
  {\it ``Unity Of Superstring Dualities,''}
  \np{438}{95}{109}, \hepth{9410167}}

\lref\GanHan{O.J. Ganor and A. Hanany,
  {\it ``Small $E_8$ Instantons and Tensionless Non-Critical Strings,''}
  \hepth{9602120}}

\lref\Gan{O.J. Ganor,
  {\it ``Toroidal Compactification of Heterotic 6D Non-Critical Strings
  Down to Four Dimensions,''} \hepth{9608109}}

\lref\GMS{O.J. Ganor, D.R. Morrison, N. Seiberg,
  {\it ``Branes, Calabi-Yau Spaces, and Toroidal Compactification
  of the $N=1$ Six-Dimensional $E_8$ Theory,''} \hepth{9610251}}

\lref\Rozali{M. Rozali,
  {\it ``Matrix Theory and U-duality in Seven Dimensions,''}
  \hepth{9702136}}

\lref\BRS{M. Berkooz, M. Rozali and N. Seiberg,
  {\it ``On Transverse Fivebranes in M(atrix) Theory on $T^5$,''}
  \hepth{9704089}}

\lref\IntSei{K. Intriligator and N. Seiberg,
  {\it ``Mirror Symmetry in Three Dimensional Gauge Theories,''}
  \hepth{9607207}}

\lref\SWQCD{N. Seiberg and E. Witten,
  {\it ``Monopoles, Duality and Chiral Symmetry Breaking in $N=2$
  Supersymmetric QCD,''} \hepth{9408099}}

\lref\BDS{T. Banks, M.R. Douglas and N. Seiberg,
  {\it ``Probing F-theory With Branes,''}
  \hepth{9605199}}

\lref\SeiIRD{N. Seiberg,
  {\it ``IR Dynamics on Branes and Space-Time Geometry,''}
  \hepth{9606017}}

\lref\SWGDC{N. Seiberg and E. Witten,
  {\it ``Gauge Dynamics And Compactification To Three Dimensions,''}
  \hepth{9607163}}

\lref\WitBR{E. Witten,
  {\it ``Solutions Of Four-Dimensional Field Theories Via M Theory,''}
  \hepth{9703166}}

\lref\ABKSS{O. Aharony, M. Berkooz, S. Kachru, N. Seiberg and
  E. Silverstein,
  {\it ``M(atrix) description of $(2,0)$ theories,''} \hepth{9707079}}

\lref\Lowe{D. Lowe,
  {\it ``$E_8 \times E_8$ Small Instantons in Matrix Theory,''}
  \hepth{9709015}}

\lref\BCD{D. Berenstein, R. Corrado, and J. Distler,
  {\it ``On the Moduli Spaces of M(atrix)-Theory Compactifications,''}
  \hepth{9704087}}

\lref\Lambert{N.D. Lambert,
  {\it ``Quantizing the $(4,0)$ Supersymmetric ADHM Sigma-Model,''}
  \np{460}{96}{221}, \hepth{9508039}}

\lref\GRT{O.J. Ganor, S. Ramgoolam and W. Taylor IV,
  {\it ``Branes, Fluxes and Duality in M(atrix)-Theory,''} \hepth{9611202}}

\lref\GanSet{O.J. Ganor and S. Sethi, to appear.}

\lref\Imamura{Y. Imamura,
  {\it ``A Comment on Fundamental Strings in M(atrix) Theory,''}
  \hepth{9703077}}

\lref\GopRam{R. Gopakumar and S. Ramgoolam,
  {\it ``Scattering of zero branes off elementary strings in Matrix theory,''}
  \hepth{9708022}}

\lref\BluInt{J.D. Blum and K. Intriligator,
  {\it ``New Phases of String Theory and 6d RG Fixed Points via Branes at
  Orbifold Singularities,''} \hepth{9705044}}

\lref\IntNEW{K. Intriligator,
  {\it ``New String Theories in Six Dimensions via Branes at Orbifold
  Singularities,''} \hepth{9708117}}

\lref\BSVtop{M. Bershadsky, C. Vafa, V. Sadov,
  {\it ``D-Branes And Topological Field Theories,''}
  HUTP-95-A047, \hepth{9511222}}
\lref\VafIOD{C. Vafa,
  {\it ``Instantons on D-branes,''} \hepth{9512078}}

\lref\SenHBP{A. Sen,
{\it ``A Note on Marginally Stable Bound States in Type II String Theory,''}
MRI/PHY/23-95, hep-th/95102229}

\lref\VafGAS{C. Vafa,
   {\it ``Gas of D-Branes and Hagedorn Density of BPS States,''} HUTP-95/A042,
   hep-th/9511088}

\lref\FMW{R. Friedman, J. Morgan, and E. Witten,
  {\it ``Vector Bundles And F Theory,''} \hepth{9701162}}

\lref\BerSad{M. Bershadsky and V. Sadov,
  {\it ``F-Theory on $K3\times K3$ and Instantons on 7-branes,''}
  \hepth{9703194}}

\lref\WitADH{E. Witten,
  {\it ``Sigma Models And The ADHM Construction Of Instantons,''}
  \hepth{9410052}}

\lref\BerCor{D. Berenstein and R. Corrado,
  {\it ``M(atrix)-Theory in Various Dimensions,''} \hepth{9702108}}

\lref\DVV{R. Dijkgraaf, E. Verlinde, H. Verlinde,
  {\it ``Matrix String Theory,''} \hepth{9703030}}

\lref\DVVF{R. Dijkgraaf, E. Verlinde, H. Verlinde,
  {\it ``5D Black Holes and Matrix Strings,''} \hepth{9704018}}

\lref\BluHar{J. Blum and J.A. Harvey,
  {\it ``Anomaly Inflow for Gauge Defects,''}
   \np{416}{94}{119},\hepth{9310035}}

\lref\WitFBR{E. Witten,
  {\it ``Five-Brane Effective Action In M-Theory,''}
  \hepth{9610234}}

\lref\HorWit{P. Horava and E. Witten,
  {\it ``Heterotic and Type I String Dynamics from
  Eleven Dimensions,''} preprint IASSNS-HEP-95-86, \hepth{9510209}.}

\lref\DLM{M. Duff, J. Liu, and R. Minasian, 
   Nucl. Phys. {\bf B452} (1995) 261.}

\lref\TowCI{P. K. Townsend, Phys. Lett. {\bf B350} (1995) 184.}

\lref\WitSD{E. Witten, Nucl. Phys. {\bf B443} (1995) 85.}

\lref\KSAB{O. Aharony, M. Berkooz, S. Kachru and E. Silverstein,
  {\it ``M(atrix) description of $(1,0)$ theories,''} \hepth{9709118}}

\lref\Lowe{D. Lowe,
  {\it ``$E_8 \times E_8$ Small Instantons in Matrix Theory,''}
  \hepth{9709015}}

\lref\SusTDL{L. Susskind,
  {\it ``T-Duality in M(atrix) Theory and S Duality in Field Theory,''}
  \hepth{9611164}}

\lref\BK{I. Brunner and A. Karch,
  {\it ``Matrix Description of M-theory on $T^6$,''} \hepth{9707259}}

\lref\BanSei{T. Banks and N. Seiberg,
  {\it ``Strings from Matrices,''} \hepth{9702187}}

\lref\Motl{L. Motl,
  {\it ``Proposals on nonperturbative superstring interactions,''}
  \hepth{9701025}}

\lref\Mart{E. Martinec,
  {\it ``Matrix theory and N=(2,1) Strings,''} \hepth{9706194}}

\lref\MotlQ{L. Motl,
  {\it ``Quaternions and M(atrix) theory in spaces with boundaries,''}
  \hepth{9612198}}

\lref\Town{P.K. Townsend,
  {\it ``M(embrane) theory on $T^9$,''} \hepth{9708034}}

% ===================================================================== %
% Introduction
% ===================================================================== %
\newsec{Introduction}

One of the standing problems in M(atrix)-theory \refs{\BFSS,\SusANO}
is to understand the compactification of M-theory
on compact manifolds.
Toroidal compactification on $\MT{d}$ for $d\le 3$ has been defined 
as $U(N)$ SYM on a dual $\MHT{d}$ \refs{\BFSS,\Wati}
(for more discussion on SYM and toroidal compactifications of 
M-theory see
\refs{\SusTDL,\GRT,\Motl,\BerCor,\BanSei,\DVV,\FHRS,\BCD,\Mart,\EGKR,\Town}).
The model for $\MT{4}$ has been studied in \Rozali\ and in \BRS\ 
and is given by the $(2,0)$ theory in 5+1D.
The ``state of the art'' in toroidal compactification is M-theory
on $\MT{5}$ \SeiVBR.
The moduli space of M-theory on $\MT{5}$ is 
$$
{\cal M}_5 = SO(5,5,\BZ)\backslash SO(5,5,\BR) / (SO(5)\times SO(5))
$$
and the theory defined in \SeiVBR\ can be thought of as a
Hilbert space with an unknown 0+1D 
Hamiltonian which depends on
25 external parameters, namely the point in ${\cal M}_5$.
At certain limiting points in ${\cal M}_5$ the 
spectrum of the Hamiltonian can be interpreted as a spectrum
of a 5+1D field theory. However, unlike ordinary compactified 
field theories there are two different limits in ${\cal M}_5$
for which the spectrum looks like a 5+1D decompactified theory.
In one limit the supersymmetry is $(2,0)$ and in the other it is 
$(1,1)$.
Moreover, in appropriate limits in ${\cal M}_5$ one recovers
also the previous theories for $\MT{d}$ with $d< 5$.

The model for M-theory on $\MT{6}$ remains elusive.
Some suggestions have been studied in \refs{\LMS,\BK,\HL} but
a large class of possibilities has been ruled out in \SeiSet\
because one cannot decouple a subset of degrees of freedom of
M-theory as in \SeiVBR.

The purpose of these notes is to explore the assumption
that there exists a Hilbert space and Hamiltonian with 16
supersymmetries which depends on parameters in 
$$
{\cal M}_6 = E_{6(6)}(\BZ)\backslash E_{6(6)}(\BR) / Sp(4),
$$
the moduli space of M-theory on $\MT{6}$ \HulTow.
For $q\in {\cal M}_6$, let us call the resulting  theory $X(q)$.

The questions that we will ask are:

\item{a.}
What is the maximal number of possible uncompactified dimensions
of $X$?
\item{b.}
What kind of extended objects does the non-compact $X$ accomodate
and what is the low energy description of these?
\item{c.}
What is the M(atrix) description of $X$?

To answer the first question,
we will use the tools developed in \WitVAR\ for the study
of degenerations of type-II string theory.
The $E_{6(6)}(\BZ)$ U-duality will then imply the existence of
extended brane-like objects in the maximally decompactified $X$.

To construct a M(atrix) model of $X$ (which we denote by $MX$)
one can try to compactify it on
a small $\MS{1}$ and find an appropriate description for multiple
Kaluza-Klein states. Using the $E_{6(6)}$ duality of $X$ we can
map the KK states to extended branes in $X$.
However, the theory on the branes will not decouple from the 6+1D bulk.
Nevertheless, we can still ask questions similar to (a)-(c) about
$MX$ and about $M^2 X$ -- the M(atrix)-model for $MX$.
 Altogether, we conjecture that a 
${\rm M}^3({\rm atrix})$
 model for M-theory on $\MT{6}$ (i.e. a M(atrix) model
for the M(atrix)-model of the M(atrix) model) is given by a 1+1D
theory with $\SUSY{(0,4)}$ supersymmetry.
We will suggest only the $p_\Vert = 1$ version of $M^2 X$ (and only
for $p_\Vert = 1$ versions of $MX$ and $X$) where $p_\Vert$ is the
longitudinal momentum in the light-like direction  (see \SusANO).

These notes are organized as follows.
Section (2) is a review of the theories involved in compactification
of M(atrix) theory on $\MT{4}$ and $\MT{5}$.
These are the $(2,0)$ theories in 5+1D (which we denote $T(N)$)
and the new 5+1D massive theory (which we denote $S(N)$).
In section (3) we review the tools of \WitVAR\ for analyzing 
degenerations and, using results of \BCD\ and \EGKR,
we apply them to the case at hand.
We find that the maximal degeneration is a 6+1D theory and we argue
that it could have a local energy momentum tensor.
In section (4) we explain the peculiar behaviour of the theory
under U-duality which involves a rescaling of units.
We explain how the 5+1D theory of \SeiVBR\ and its spectrum arise
in limiting cases.
In section (5) we study the low-energy descriptions of
long extended objects in
6+1D. These are 2-branes and 5-branes. We argue that when two
5-branes coincide an interacting 
 theory with $(1,0)$ SUSY in 5+1D and a low-energy with tensor
multiplets appears. Similarly, we study two coinciding 2-branes
and argue that a 2-brane can end on a 5-brane.
We argue that reparameterization anomalies could be canceled.
In section (6) we study the required properties of a M(atrix)-model
for $X$ and for its M(atrix)-model.

% ===================================================================== %
% Review
% ===================================================================== %
\newsec{Review}

There are two kinds of SUSY algebras with 16 supersymmetries in 5+1D.
The spinor representations $\rep{4}$ and $\rep{4'}$ of $SO(5,1)$ are
pseudo-real, each with 8 real components.
The first kind of SUSY algebra has SUSY charges
one in the $\rep{4}$ and one in the $\rep{4'}$
and is referred to as $(1,1)$. $U(1)$
Super-Yang-Mills theory in 5+1D has this kind of symmetry.
The other kind is $(2,0)$ and has two SUSY charges in the 
$\rep{4}$ of $SO(5,1)$.
A free tensor multiplet is a realization of such a SUSY algebra.
It comprises of an anti-self-dual tensor
field-strength $G_{\u\v\tau}^{(-)}$ (satisfying 
$G_{\u\v\tau}^{(-)} = 
-{\epsilon_{\u\v\tau}}^{\u'\v'\tau'} G_{\u\v\tau}^{(-)}$
and $\px{\lbrack \sigma}G_{\u\v\tau\rbrack}^{(-)} = 0$) together with
five real scalars $\Phi^I$ and two fermionic spinor fields.
In the introduction we denoted this free theory by $T(1)$.

The two examples above, $U(1)$ and $T(1)$ are free.

What about interacting theories in 5+1D?
Generalizing $U(1)$ to $U(N)$ Yang-Mills is possible at the classical
level, but this leads to a non-renormalizable field theory.
However, a recent conjecture \SeiVBR\ has been put forward for
an interacting 5+1D theory which at low-energies can be approximated
by 5+1D $U(N)$ SYM (which is a free theory of $N^2$ gluons with
IR-irrelevant cubic and quartic interactions).
To be more precise, this new theory has two kinds of low-energy
limits.
Taking the limit in one direction of the parameter space
one obtains a low-energy of massless vector multiplets.
In another limit, one obtains massless tensor multiplets.

We will review the arguments of \SeiVBR\ momentarily, but as a
preliminary we will discuss the generalization of $T(1)$ to
an interacting theory $T(N)$.

% --------------------------------------------------------------------- %
% Review of T(N)
% --------------------------------------------------------------------- %
\subsec{Review of $T(N)$}

The interacting generalization of $T(1)$ was discovered in \WitCOM\
by studying type-IIB on an $A_N$ singularity.
Another realization of the theory is the low-energy description of
$N$ coincident 5-branes of M-theory \StrOPN.
$T(N)$ has a moduli space corresponding to separation of the
5-branes. The low-energy at a generic point in the moduli space is
described by $N$ free tensor multiplets. The 5 scalar fields of each
multiplet parameterize the moduli space. In uncompactified $\MR{5,1}$,
$T(N)$ has no BPS particles but it has BPS 1-branes. They are charged
under the field strength of the tensor multiplet and their tension
is given by the square root of the sum of squares of the scalars 
which are the superpartners of the corresponding 3-form field
strength \SWSIXD.
In the 5-brane pictures these 1-branes are membranes with a boundary
on a pair of 5-branes \StrOPN.
One of the discoveries in \WitCOM\ was that when $T(N)$ is compactified
on $\MT{2}$ and when the size of the $\MT{2}$ is much smaller than
the scale of $T(N)$ (derived from the VEVs of the scalars) the
resulting low-energy description in the uncompactified 3+1 dimensions
is 3+1D $U(N)$.
Thus $U(N)$ can be derived as a limit of $T(N)$.
% --------------------------------------------------------------------- %
% Review of S(N)
% --------------------------------------------------------------------- %
\subsec{Review of $S(N)$}

The theory $S(N)$ is defined \SeiVBR\ as the degrees of freedom
which describe
$N$ coincident NS5-branes in type-IIB in the limit $\lam_s \rightarrow 0$
while keeping $M_s$ (or equivalently, $\a'$) fixed. More precisely,
this limit is described by a tower of perturbative non-interacting
string states which live in the 9+1D bulk 
supplemented by interacting degrees
of freedom on the 5+1D NS5-brane world-volume \SeiVBR.
The uncompactified $S(N)$ theory has a low-energy description of $N$ 
free vector-multiplets on $\MR{5,1}$.
The $S(N)$ theories are not scale invariant but have
a scale $M_s$.
 $S(N)$ has a moduli space parameterized by the $4N$ scalars
of the vector-multiplets.
 It was argued in \SeiVBR\ that $S(N)$ has BPS 1-brane 
excitations which can be thought of as bound states
 of elementary strings and the NS5-branes.
By this we mean that they are charged under the bulk NS-NS
2-form when one turns on $\Ol{\lam_s}$ corrections.
These excitations have tension $M_s^2$.

All that has been said above is about $S(N)$ on the non-compact $\MR{5,1}$.
One of the fascinating properties of $S(N)$ is that the space-time
interpretation of the theory is not unique!
Once one compactifies the theory on $\MT{5}$ the theory  has  an
$SO(5,5,\BZ)$ T-duality which means that the Hamiltonian and Hilbert 
space describing $S(N)$ on $\MT{5}$ can be mapped in a 1-to-1 way
to the Hamiltonian and Hilbert space of $S(N)$ on another $\MT{5}$
related by an $SO(5,5,\BZ)$ T-duality transformation to the original one.
In this map, momentum states become extended states.
The existence of such a theory has been previously conjectured
in \refs{\DVVQ,\DVVS} where the relation with the NS5-brane has also
been suspected. The limit of $\lam_s\rightarrow 0$ and the 
decoupling argument in \SeiVBR\ give a concrete realization of
such a theory.

By studying the limit of $S(N)$ on a $\MT{5}$ with all sides shrunk to zero
one obtains a theory which can be given another space-time interpretation
with a different SUSY structure.
This theory is realized as the limit of $N$ NS5-branes of type-IIA
when $\lam_s\rightarrow 0$ keeping $M_s$ fixed \SeiVBR.
This theory has a low-energy description of $N$ 
free tensor multiplets on $\MR{5,1}$.
The moduli space is parameterized by the $5N$ scalars
of these tensor-multiplets.
However, $N$ of the scalars have a compact moduli
space $\MS{1}$ of radius $M_s^2$ \BRS. The full moduli space is
$(\MR{4}\times\MS{1})^N/S_N$.

% --------------------------------------------------------------------- %
% Low-energy limits
% --------------------------------------------------------------------- %
\subsec{Locality and low-energy limits of $S(N)$}

We will adopt the following point of view in this paper.
$S(N)$ compactified on $\MT{5}$ is to be thought of as some unknown
Hamiltonian $H$ acting on some unknown Hilbert space $\Hil$.
The Hamiltonian depends on the geometric parameters of the $\MT{5}$,
namely the 5 radii $R_1,\dots,R_5$, the angles and the NS-NS 
background B-fields.
They span the  space
$$
\MSO_5 = SO(5,5,\BZ)\backslash SO(5,5,\BR) / SO(5)\times SO(5).
$$
These parameters are to be thought of as {\it external} constant
parameters.
At a generic point of $\MSO$ the theory is completely compactified
and, like any compactified field theory, $H$ has a discrete spectrum.
At various limits of the parameters of $\MSO_5$ one can discover that the
low-energy spectrum of $H$ can be approximated by a field theory
on a very large space.
There are two such low-energy limits.
One limit is to take all radii $R_i\rightarrow\infty$.
In this limit the low-energy excitations
correspond to $N$ free tensor-multiplets on $\MR{5,1}$.
(To be more precise, the Hilbert space is approximated by
super-selection sectors parameterized by the VEVs of the 
tensor fields.)

There is another limit of parameters, $R_1\rightarrow 0$ and 
$R_2,\dots R_5\rightarrow\infty$,
in which the low-energy excitations correspond to $N$ free
vector-multiplets.
For generic values of the parameters there is no unique space-time
interpretation but this doesn't mean that there is no local
interpretation at all. On the contrary, there are probably many good
local descriptions of the theory.
(By locality, I mean that there is a set of local operators,
the Hamiltonian can be written as an integral of a local operator
and the commutation relations are polynomial in derivatives).

However, there is no ``canonical'' local description.
There is an $SO(5,5,\BZ)$ isometry of the Hilbert space which acts
on the Hamiltonian as a T-duality on the parameters of the $\MT{5}$.
Local operators in one description
are not local in another description.

Over the rest of this paper we will assume that there are good local
descriptions. We must mention, however, one reason that might lead
one to doubt that statement.
The theory has a Hagedorn-like spectrum of states (this can be seen
by counting BPS states of wound strings \EGKR)
there might be a problem
in defining short-distance Operator-Product-Expansions for
distances shorter than the inverse of the Hagedorn temperature $M_s$.

% ===================================================================== %
% Section (): Maximal decompactification}
% ===================================================================== %
\newsec{Maximal decompactification limits}

The purpose of this paper is to explore a possible M(atrix) description
for M-theory on $\MT{6}$.
We start with the assumption that there exists a Hilbert
space with a Hamiltonian $H$ depending on external parameters in
$$
\MSO_6 = E_{6(6)}(\BZ) \backslash E_{6(6)}(\BR) / Sp(4)
$$
and having 16 supersymmetries.
Our first question will be what is the limit in $\MSO$
for which a low-energy
description with a maximum number of dimensions is attained.

The way to settle this question is to follow an analysis similar
to that given in \WitVAR.
There, the BPS formula for BPS excitations was used to determine
the degenerations with a maximal number of states becoming light.
These states were then interpreted as KK states.
Let us recall the details of the analysis in \WitVAR.
For a moduli space of $G/K$ with $G=E_{d(d)}$ and $K$ a maximal
compact subgroup, the central charge $Z^{ij}$ transforms
in a certain representation $\rep{R}_K$ of $K$.
Denote the space of $Z^{ij}$ as $W$.
The space of physical charges is in a representation $\rep{R}_G$
of $G$. Denote it by $V$. 
The BPS mass formula is (in $(11-d)$-dimensional Einstein units):
$$
M_E = \left\| Z\right\| = \left\| T g^{-1}\psi \right\|
$$
for $\psi$ a vector of charges $g\in G/K$
and $T$ some fixed map $T:V\rightarrow W$.

The representations for different dimensions are (see \BCD\ and \EGKR):
\bigskip
\begintable
$d$          |     4       |    5                |      6     |  7   \elt
$G$          |     $SL(5)$ |  $SO(5,5)$          | $E_6$      |  $E_7$ \elt
$\rep{R}_G$  |  $\rep{10}$ |   $\rep{16}$        | $\rep{27}$ | $\rep{56}$ \elt
$K$          |    $Sp(2)$  | $Sp(2)\times Sp(2)$ | $Sp(4)$    | $SU(8)$ \elt
$\rep{R}_K$  |  $\rep{10}$ | $\rep{16}$          | $\rep{27}$ | $\rep{28}$ 
\endtable
\bigskip

In our case, we wish to determine the maximal number of light states
of the {\it M(atrix)-model} and not of space time itself.
The difference will be that the  representations
$\rep{R}_K$-s and $\rep{R}_G$-s change.
One way to think about it is that the space-time KK states
are electric and magnetic fluxes in the M(atrix)-model.
The representations for the KK states of the M(atrix) model are
(see \BCD\ and \EGKR):
\bigskip
\begintable
$d$          |     4       |    5                |      6      \elt
$G$          |     $SL(5)$ |  $SO(5,5)$          | $E_6$       \elt
$\rep{R}_G$  |  $\rep{5}$  |   $\rep{10}$        | $\rep{27}'$  \elt
$K$          |    $Sp(2)$  | $Sp(2)\times Sp(2)$ | $Sp(4)$     \elt
$\rep{R}_K$  |  $\rep{5}$  | $\rep{10}$          | $\rep{27}'$
\endtable
\bigskip
These are the same as the representations for longitudinal branes
in M-theory \refs{\Imamura,\GopRam}.

Now we proceed to analyze the degeneration with a maximal number
of massless particles.
It is sufficient to restrict to diagonal matrices $g\in G$.
% --------------------------------------------------------------------- %
% $\MT{4}$
% --------------------------------------------------------------------- %
\subsec{Calculation for $\MT{4}$}

For M-theory on $\MT{4}$, the generic diagonal
matrix of $\rep{5'}$ has eigenvalues
\eqn\esix{
V, r_1^{-1}, r_2^{-1}, r_3^{-1}, r_4^{-1},
}
where $V = \prod_1^4 r_i$.
For $n_1,\dots,n_5\in\BZ$ there are BPS states with squared masses
$$
\sum_1^4 (n_i r_i^{-1})^2 + (n_5 V)^2.
$$
These states are interpreted as the KK states of a theory in 5+1D
compactified on $\MT{5}$ with radii $\u r_1,\dots, \u r_4, \u V^{-1}$
where $\u$ is a parameter that stands for
a possible rescaling of the units of energy.
Indeed, we know that we expect the $(2,0)$ 5+1D theory which
is conformally invariant \refs{\Rozali,\BRS}.

% --------------------------------------------------------------------- %
% $\MT{5}$
% --------------------------------------------------------------------- %
\subsec{Calculation for $\MT{5}$}

For M-theory on $\MT{5}$, the generic diagonal
matrix of $\rep{10}$ has eigenvalues
\eqn\esix{\eqalign{
r_i^{-1},&\qquad i=1\dots 5,\cr
r_i,&\qquad i=1\dots 5,\cr
}}
For every $n_1,\dots,n_5$ there exists a BPS state with mass squared
\eqn\thex{
\sum_1^5 (n_i r_i^{-1})^2.
}
There are also BPS states with masses squared
$$
(n_1 r_1)^2 + \sum_2^5 (n_i r_i^{-1})^2,\,
(n_1 r_1)^2 + (n_2 r_2)^2 + \sum_3^5 (n_i r_i^{-1})^2,\,\cdots,\,
\sum_1^5 (n_i r_i)^2.
$$
We see that there are several low-energy limits.
For example, taking $r_i\rightarrow\infty$
we find 5 decompactified dimensions.
The states in \thex\ are KK states of this theory.
There are also BPS states with masses squared
$$
(r_1)^2 + \sum_2^5 (n_i r_i^{-1})^2.
$$
The existence of these states can be deduced entirely from U-duality.
The fact that only directions $2\dots  5$ appear in
$\sum_2^5 (n_i r_i^{-1})^2$ indicates that these states can be given
momentum only in 4 directions out of the five (and still be BPS).
This gives them up as extended strings which
 can have momenta in 4 transverse directions.
 Their tension is constant and is set to 1.
The fact that the tension is set (and does not depend on $r_i$)
is consistent with a local physics.
In fact, the existence of a local energy momentum tensor
can also be deduced from the construction of \SeiVBR.
Turning on $\Ol{\lam}$ corrections to the system of an NS5-brane
in weakly coupled type-IIA one learns that for a consistent
coupling to the bulk gravity there has to be a local energy momentum
tensor on the 5-brane.\foot{
I learned that this argument has also been given by L. Susskind.}

Another low energy limit is obtained when $r_i\rightarrow 0$
and the analysis is similar.

% --------------------------------------------------------------------- %
% $\MT{6}$
% --------------------------------------------------------------------- %
\subsec{Calculation for $\MT{6}$}

For M-theory on $\MT{6}$, the generic diagonal
matrix of $\rep{27'}$ has eigenvalues
\eqn\esix{\eqalign{
V^{-1/3}r_i^{-1},&\qquad i=1\dots 6,\cr
V^{-1/3} r_i r_j,&\qquad 1\le i<j\le 6,\cr
V^{-1/3} \prod_{j\ne i} r_j &\qquad i=1\dots 6,\cr
}}
where $V = \prod_1^6 r_i$.
The parameters $r_i$ have been chosen so 
as to exhibit the subgroup $SL(6)\times SL(2)$ of $E_{6(6)}$.
The six values $r_i/V^{1/6}$ form an $SL(6)$ matrix and 
the remaining parameter $V$ is related to the $SL(2)$ generator.\foot{
I am grateful to E. Witten for pointing this out.}
By applying $E_{6(6)}(\BZ)$ we see that all combinations
$$
V^{-1/3}\sqrt{\sum_1^6 (n_i r_i^{-1})^2},\qquad n_i\in \BZ
$$
exist as masses of BPS states.
There are also states with masses
$$
V^{-1/3} \sqrt{(r_i r_j)^2 + \sum_{k\ne i,j} (n_k r_k^{-1})^2},
$$
and states with masses
$$
V^{-1/3} \sqrt{(\prod_{j\ne i} r_j)^2 + (n_i r_i^{-1})^2}.
$$
We will interpret these formulae in the next section.
% ===================================================================== %
% Section (4): Limits of the theory
% ===================================================================== %
\newsec{The 6+1D limit and its compactification}

In this section we will argue that $X$ has a 6+1D limit.
We will study its properties and
recover the properties of the  5+1D theory of \SeiVBR\ from
compactification on $\MS{1}$.

The assumption that a M(atrix) model for M-theory on $\MT{6}$
exists lead us to the conclusion that there is a 0+1D Hamiltonian
$H(r_1,\dots,r_6)$ where $r_i$ are just external parameters for now.
From this Hamiltonian we can define a new Hamiltonian
\eqn\whh{
\wH (r_1,\dots,r_6) = V^{1/3} H(r_1,\dots, r_6),\qquad
V = r_1\cdots r_6.
}
From the BPS formula we learn that the BPS particles of $\wH$
have masses
\eqn\bpsma{\eqalign{
{m_{\{n_k\}}}^2 &= \sum_1^6 (n_i r_i^{-1})^2,\qquad n_i\in \BZ,\cr
{m_{i,j,\{n_k\}}}^2 &= (r_i r_j)^2 + \sum_{k\ne i,j} (n_k r_k^{-1})^2,\cr
{m_{i,n_i}}^2 &= (\prod_{j\ne i} r_j)^2 + (n_i r_i^{-1})^2.
}}
All these states are related by the conjectured U-duality of the
theory.
In the limit $r_i\rightarrow\infty$ the first line constitutes
the lightest particles.
We see that the first line is consistent with an interpretation
of momentum states of a Lorenz invariant field theory on $\MT{6}$.
The second line represents objects which can have momentum only
in 4 directions. Thus, we must interpret them as 2-branes.
We see that their tension is constant.
In fact, the rescaling factor $V^{-1/3}$ in \whh\ was chosen such
that the tension of the 2-branes will be independent of the $r_i$-s.
The third line of \bpsma\ represents objects which can have momentum
only in one transverse direction. They must be 5-branes and 
in the units \whh, the tension of these is constant.

Thus, formula \bpsma\ is consistent with a Lorenz invariant
6+1D interpretation.

The 6+1D low-energy of the theory would then have to be
$N^2$ free vector multiplets since there are no interacting 
conformal theories in 6+1D \SeiNTS.

We are studying the $N=1$ version of the theory 
so the low-energy must be $U(1)$ 6+1D Yang-Mills.
Let us see what the coupling constant of the Yang-Mills would be.
The M(atrix) conjecture requires us to identify $V^{-1/3} r_i^{-1}$
with the tension of a longitudinal object in M-theory on $\MT{6}$.
The tension must be measured in Einstein units in 4+1D and so 
we find \foot{This formula was calculated together with N. Seiberg and
S. Sethi.}
$$
R_i = V^{-2/9} r_i
$$
where $R_i$ are the radii of the $\MT{6}$ in 10+1D units.
The energy of an electric flux in direction $i$ is (see also \EGKR):
$$
{{g^2 r_i^2} \over {V}}.
$$
This is to be identified with a square of the mass of a
KK state of M-theory on $\MT{6}$
and we find that $g$ is a constant as well.

% --------------------------------------------------------------------- %
% Another IR limit
% --------------------------------------------------------------------- %
\subsec{Another IR limit}

Like the theory of \SeiVBR, $X$ has more than one IR limit.
However, a novel feature of $X$ is that to describe the other IR limits
we must change the units of energy.

Let us take the limit $r_i\rightarrow 0$ in \bpsma.
In this limit, the lightest BPS states will be the wrapped 5-branes
so we will have to identify them with KK states of the new IR limit.
Thus we will write
$$
R_i^{-1} = {V\over {r_i}},
$$
where the RHS is the mass of a wrapped 5-brane transverse to the $i$-th 
direction. $R_i$ are the radii of the new IR limit.
However, the 2-branes now have tension
$$
r_i r_j = (\prod R_i)^{-2/5} R_i R_j.
$$
Their tension is not constant and this is not a good local description
of physics.
The resolution is that there was no reason why the new IR limit should
keep the same unit of time as before. After all, the energy direction
is not in a Lorenz multiplet with the 5-brane charge. It is only
in a multiplet with a momentum charge.
Thus, we choose to rescale the energy by a factor of $V^{-2}$
and define
$$
R_i' = V^{-3} r_i.
$$
The new Hamiltonian
$$
\wH(R'_1,\dots, R'_6) = V^{-2} H(r_1,\dots, r_6)
$$
defines a local 6+1D theory compactified on radii $R'_i$.

% --------------------------------------------------------------------- %
% Recovery of lower dimensional theories
% --------------------------------------------------------------------- %
\subsec{Recovery of lower dimensional theories}

We can recover the $SO(5,5)$ theory described in
\SeiVBR\ by compactifying the 6+1D theory on a small $r_6$.

We first find a theory with a low-energy of 5+1D $U(1)$ SYM and coupling
constant
$$
{1\over {g^2}} \sim r_6
$$
since the 6+1D coupling constant was 1.
This theory has string states (which are wrapped membranes) of tension
proportional to $r_6$. It also has membranes with a constant tension,
4-branes with a tension of $r_6$ and 5-branes with a constant tension.
The standard analysis shows that as $r_6\rightarrow 0$ the strings
are the lowest lying states.
We should now change units (both time and space) so that
the coupling constant will be 1 in the new units. This also sets
the tension of strings to 1 and all the higher states are pushed
to infinite energy as $r_6\rightarrow 0$.

% ===================================================================== %
% Section (5): Extended objects
% ===================================================================== %
\newsec{Extended objects}

We have seen in the previous section that there must be
a limit in moduli space for which the spectrum can be
interpreted as a 6+1D local theory.
The 6+1D low-energy of that theory would then have to be
$N^2$ free vector multiplets since there are no interacting 
conformal theories in 6+1D.

However, we can ask other ``low-energy questions'' for which
the answer is nontrivial.
We have seen that the theory has extended BPS 5-branes and extended
BPS 2-branes.
Note that because the 5-branes are not charged in this theory
there are no IR divergences even when there is only one transverse
direction.

% --------------------------------------------------------------------- %
% 5-branes
% --------------------------------------------------------------------- %
\subsec{5-branes}

What is the low-energy description of a theory with an extended
5-brane?

If the theory is local, the description has to be in terms of
extra degrees of freedom ``living'' on the 5-brane and interacting
with the bulk low-energy fields.
The 5+1D theory  has $\SUSY{(1,0)}$ supersymmetry and
its moduli space is $\BR$, the motion in the single transverse
direction. Thus, it can only be a free tensor multiplet.

What happens when two 5-branes coincide?
We expect to find a 5+1D conformal theory with a moduli space
of $(\BR)^2/\BZ_2$ where the $\BZ_2$ exchanges the two $\BR$-s.
After compactification on $\MT{3}$ this theory should have a moduli
space of $(\MT{4})^2/\BZ_2$. This is because when compactified on
another $\MT{2}$,
 the system of two 5-branes can be mapped by the $E_{6(6)}(\BZ)$
U-duality to a system of two KK states
and two KK states have $(\MT{6})^2/\BZ_2$ as their ``moduli'' space.
Let us first rule out a few possibilities for such a theory.
Various 5+1D theories with $(1,0)$ SUSY and tensor multiplets in
the low-energy have been studied in \refs{\BluInt,\IntNEW}.
These theories contained in general also hyper-multiplets in the 
low-energy and so we believe that the theory we are looking for is
not in that list.
Another possibility is to separate the tensor multiplet corresponding
to the center of mass motion of the two 5-branes
and be left with a moduli space of $\BR/\BZ_2$. 
There exists a 5+1D theory with a low-energy comprising of 
a single tensor multiplet with moduli space $\BR/\BZ_2$.
This is the theory associated with small $E_8$ instantons
\refs{\GanHan,\SWSIXD}.
However, this theory cannot be the one we are looking for either.
In fact, the mere assumption that the moduli space in 5+1D
is $\BR/\BZ_2$ strongly suggests that the theory has an $E_8$ 
symmetry. The line of reasoning is as follows \GMS.
When one compactifies such a theory down to 3+1D on a torus
one finds an $\SUSY{2}$ theory with a vector multiplet.
The moduli space at infinity is determined from the fact
that we reduced a 5+1D tensor multiplet. This restricts
the form of the Seiberg-Witten curve.
The $H_2(\BZ)$ cohomology of the total space (moduli space together
with the elliptic fibers)
is related to global symmetries of the theory \SWQCD.
It can be shown that $H_2(\BZ)$ contains an $E_8$ lattice
\refs{\Gan,\GMS}.
It is unreasonable to expect a {\it global} $E_8$ symmetry from
the system just described.
M-theory on $\MT{6}$ doesn't have an $E_8$ global 
symmetry.
Furthermore, the BPS excitations of the small $E_8$ instanton theory
are strings which carry an $E_8$ chiral current algebra.
In our case, these strings should be identified with 2-branes
with boundaries on the two 5-branes (we will elaborate on this
later on) and it is not clear from where the $E_8$ chiral current
algebra would come.
We see that the $E_8$ theory has to be ruled out.
In fact, after compactification on $\MT{3}$ the moduli space
has to be $(\MT{4})^2/\BZ_2$ and not include half a $K_3$.

Thus, we conclude that if the theory $X$ exists, there must be a
new 5+1D theory with $(1,0)$ SUSY and a low energy of $k$ 
tensor multiplets with moduli space $(\BR^k)/ S_k$.
For the moment we will assume that such theories exist.

% --------------------------------------------------------------------- %
% 2-branes
% --------------------------------------------------------------------- %
\subsec{2-branes}

An extended 2-brane has 4 transverse directions and it is
natural to expect that the position in these directions
is related to the VEVs of 4 scalars.
When $k$ 2-branes coincide
we find a 2+1D theory with a moduli space of
$(\MR{4})^k/S_k$.
Let us restrict to the case of a single 2-brane.
The 4 scalars could be super-partners 
of either a vector multiplet or a hypermultiplet
of $\SUSY{4}$ in 2+1D.
Locally these two possibilities are identical
(2+1D mirror symmetry \IntSei\ replaces vector multiplets with
hypermultiplets).
Globally, there is a difference.
If the world-volume of the 2-brane is a genus-$g$ Riemann surface,
a vector multiplet will produce $2g$ global modes (the Wilson lines).

When the directions transverse to the 2-brane are compact
we can use U-duality to map the 2-brane to a 5-brane.
Since the world-volume of a 5-brane is described by a tensor field
which upon dimensional reduction becomes a vector field, we see
that the global modes do exist.

These global modes affect
the counting of $\half$-BPS states in $X$.\foot{This
setting is very similar to a situation discussed together
with S. Sethi in a different context \GanSet.}
Let $X$ be compactified on $\MT{6}$ and
let $k$ 5-branes be in directions
$1\dots 5$ and let $k'$ units of momentum be on the 5-branes
in direction $5$. When the 5-th direction is small we can think 
of the 5-brane theory as 4+1D $U(k)$ Yang-Mills
compactified on $\MT{4}$. The $k'$ units of momentum become
instantons in $U(k)$ Yang-Mills with 8 supersymmetries.
The configuration is $\half$-BPS and the multiplicities
and quantum numbers
of states are found by quantizing this moduli space (similarly to
\refs{\SenHBP,\VafGAS,\BSVtop,\VafIOD}).
On the other hand, we can use $E_{6(6)}$ U-duality  to map
this system to a system of $k$ 2-branes wrapped on directions $1,2$
and $k'$ 2-branes wrapped on directions $3,4$.
There are two compact transverse directions.
In general the $k$ and $k'$ branes form a sort of bound state
which is a single holomorphic curve in $\MT{4}$ which wraps
$k$ times around the directions $1,2$
 and $k'$ times around the directions $3,4$.
The moduli space of instantons in $\MT{4}$ is related to the
moduli space of holomorphic curves in the dual torus
$\MHT{4}$ via the {\it spectral
curve} construction (see \BerSad\ and \FMW).
In general the moduli space of $U(k)$ instantons with instanton number
$k'$ can be constructed as the moduli space of holomorphic curves
as above (the ``spectral-curve'') but in addition we need to specify
a flat line-bundle on the curve (the ``spectral-bundle'').
Thus, we conclude that there must be a $U(1)$ vector-field
living on the 2-brane with the global Wilson line modes.

% --------------------------------------------------------------------- %
% 2-branes ending on 5-branes
% --------------------------------------------------------------------- %
\subsec{2-branes ending on 5-branes}

In M-theory 2-branes can end on 5-branes and the end-point is
a source for the tensor field-strength on the 5-branes \StrOPN.
Can a 2-brane of $X$ end on a 5-brane of $X$?
We will bring supporting evidence that it can.

Let us compactify $X$ on a very large $\MT{5}$
and put in a single 5-brane that wraps $\MT{5}$.
This is a BPS state of the Hamiltonian. Now let us put in a flux
for the anti self-dual field-strength $G_{0ij}$ of the tensor field
that lives on the 5-brane. Such a flux breaks the SUSY by another half.
Since the final state is BPS the flux $G_{0ij}$ must correspond
to a central charge. The only central charges of $X$ are those in the
$\rep{27}'$ of $Sp(4)$ so we must identify the flux-charge 
with the central charge of 2-branes.
Now, let the 5-brane be stretched in directions $1\dots 5$ and
let a 2-brane be stretched in directions $1,6$.
The 1+1D endpoint of the membrane on the 5-brane acts as a source
for $G_{01j}$ so that integrating the flux through a sphere inside
the 5-brane which surrounds the endpoint of the 2-brane
$$
\int_{\MS{3}} G_{01 j} dn^j
$$
we get on the one hand the charge of the string inside the 5-brane theory
and on the other hand we get the number of 2-branes according to the
identification of $G_{0ij}$ with the 2-brane charge.
I do not know for sure from this setting
if one unit of flux corresponds to a
single 2-brane or more.

% --------------------------------------------------------------------- %
% A 2-brane stretched between two 5-branes
% --------------------------------------------------------------------- %
\subsec{A 2-brane stretched between two 5-branes}

The 5+1D theory which describes two 5-branes has a low energy
of two $(1,0)$ tensor multiplets. It was explained in \SWSIXD\
that the scalar components of tensor multiplets are the central
charges for strings which are charged under the tensor field.
This leads one to wonder whether there actually exist BPS
strings in the 5+1D theory that are charged under the difference
of the two tensor fields. From the discussion above it seems
that the answer is positive. 2-branes with boundaries on
both 5-branes will be charged under the difference of tensor fields
just like the analogous situation in M-theory \StrOPN.

What is the low-energy description of such a string in 5+1D?

For a single membrane stretched between two 5-branes of M-theory
the low-energy description is given by a 1+1D theory with 4 free scalars
and $\SUSY{(4,4)}$ supersymmetry. It can be thought of as 
the dimensional reduction of either a 5+1D hypermultiplet or a 5+1D
vector multiplet of $\SUSY{(1,0)}$ down to 1+1D.
In our case, a standard analysis of the unbroken SUSY charges reveals
that a single 2-brane is described by a 1+1D field theory with 4
scalars and $\SUSY{(0,4)}$ supersymmetry.

% --------------------------------------------------------------------- %
% 5-branes with a transverse circle
% --------------------------------------------------------------------- %
\subsec{5-branes with a transverse $\MS{1}$}

Let us compactify $X$ on a small circle of radius $r\ll g^{2/3}$
and let us put in $k$ 5-branes which fill the uncompactified 5+1D
directions.
At energies $E\ll r^{-1}$ the theory looks like a 5+1D theory.
The 6+1D bulk reduces to the theory of \SeiVBR\ with the energy
scale $g^{-1} r^{1/2}$.
At low-energies it looks like a 
5+1D Yang-Mills with coupling constant
$g^2/r$.
The $k$ 5-branes give $k$ tensor multiplets at low-energy but
this time the moduli space is $(\MS{1})^k/ S_k$.
This theory has $SO(5,5)$ T-duality and looks like a $\SUSY{(1,0)}$
``cousin'' of the theory of \SeiVBR.
The ``bulk'' theory does not decouple from the theory
of the $k$ 5-branes.

% --------------------------------------------------------------------- %
% Remarks on reparameterization anomalies
% --------------------------------------------------------------------- %
\subsec{Remarks on reparameterization anomalies}

The low-energy  description of the extended 5-brane as well
as the description of the 2-brane stretched between two 5-branes
are chiral theories and the issue of gravitational anomalies
arises.

Let us start with the 5-brane.
Since $X$ does not include gravity, the gravitational anomaly of the
tensor multiplet is not a problem when the 5-brane world-volume
is flat.

When the 5-brane world-volume is a curved 5+1D manifold the
anomaly under reparameterization is a problem even before coupling
to gravity. As explained in \HorWit, there is no canonical coordinate
frame for the 5+1D world-volume when it is curved
and we need to provide an expression for the action which is independent
of the particular coordinate system.

The anomaly of a tensor multiplet has been calculated in \WitFBR\
and does not vanish
(Our situation is even simpler since the normal bundle to the 5-brane
is trivial).

The present setting allows for a unique solution to the anomaly problem.
Since the 5-brane fills a co-dimension 1 oriented manifold the 5-brane
divides 6+1D space-time into disconnected regions.
For simplicity, let us assume that the 5-brane bounds a 6+1D
region $\Sigma$. Let $J_7$ be the 7-form such that
$$
d J_7 = I_8(R),
$$
where $I_8(R)$ is the 8-form from which the 6-form anomaly
is derived.
We can add a term
\eqn\jseven{
\int_\Sigma J_7(\omega),\qquad \partial\Sigma = \lbr
{\rm fivebrane\ worldvolume}\rbr
}
to the action which will cancel the anomaly.
If the ambient space were in dimension higher than 6+1D, there would not
have been a canonical $\Sigma$.
For flat space, different choices of $\Sigma$ would give the same result
since $d J_7 = I_8 = 0$,\foot{I am grateful to W. Taylor for pointing this
out.} but the form \jseven\ suggests that the anomaly
could be canceled on curved spaces as well.
There are situations for which the region $\Sigma$ cannot be
well defined, for example for a single 5-brane wrapping a face
of a $\MT{6}$. However, in this case the total homology class of all the
5-branes is a good quantum number and the relative phase
of the wave function has to be well-defined only within
the homology sector. We hope that the difference in the actions
\jseven\ for two configurations in the same class can still be
well defined, though we have not checked this.

The situation of a 2-brane ending on a 5-brane can have a 
different solution. The reparameterization anomaly is supported 
at the 1+1D boundary of the 2-brane on the 5-brane.
It is derived by descent equations from $\trp{R^2}$.
Part of the anomaly could be canceled by adding a term
$\int\omega_3$ where $\omega_3$ is the Chern-Simons
3-form such that $d\omega_3 = \trp{R^2}$ and
the integral is over the 2-brane world-volume.

The anomaly could also be canceled if there is an extra chiral matter
coming from the 1+1D boundary of the 2-brane on the 5-brane.
The fact that the 5-brane is immersed in 6+1D is again important.
If the codimension of the 5-brane in its ambient space-time were
higher than 1 we could have argued by Lorenz invariance that there
is no preferred direction ``left'' or ``right'' on the 1+1D
boundary. In our case, since the 5-brane has an orientation and 
the 6+1D bulk has its own orientation there is a distinction between
``left'' or ``right'' of the 5-brane. Since the 2-brane is oriented
as well there is also a distinction between ``left'' and ``right''
on the 1+1D boundary.
This also suggests that under a parity transformation the 5-brane
transforms into an anti-5-brane.

I do not know what is the composition of the required chiral matter.
In section (6.2) we will encounter
a related situation which will require fermions in a chiral 
representation of the transverse space but that suggestion is incomplete.

Finally, we note that the anomaly 
cannot be canceled by an inflow mechanism \BluHar,
i.e. by adding a term
\eqn\inflow{
\int_{\rm 5-brane} G_3\wdg \omega_3
}
where $G_3$ is the anti-self-dual field strength on the 5-brane
(in analogy with the $\int C_3\wdg I_8(R)$ term
of M-theory \refs{\DLM,\TowCI,\WitSD}). 
The term \inflow\ has opposite signs depending on whether the 2-brane
is on one side of the 5-brane or the other whereas the anomaly
itself does not change sign.

% ===================================================================== %
% Section (6): A Matrix model for X}
% ===================================================================== %
\newsec{A M(atrix) model for $X$}

In this section we will assume that $X$ is well defined
and study a possible M(atrix) model which defines 
$X$ compactified on $\MT{5}$.\foot{The first idea of looking for
a M(atrix) model for a M(atrix) model was presented in \MotlQ.}
Note that this is the maximal compactification that can be achieved
in the IMF. We will denote the $p_\Vert = k$
DLCQ sector of $X$ by $MX(k)$.

When $X$ is formulated on $\MT{5}\times\MR{1,1}$ part of the 
$E_{6(6)}(\BZ)$ U-duality is preserved. This part is the $SO(5,5,\BZ)$
acting as T-duality on the $\MT{5}$ (including the $B$-fields).
The M(atrix)-model for $X$ should preserve that $SO(5,5,\BZ)$
T-duality.

Another requirement is that $MX$ will have
8 supersymmetries and the fermionic zero modes should, upon
quantization, give the states of a vector-multiplet which will
become the $U(1)$ photon quantum numbers.

Finally, we expect $MX$ 
fluxes which will correspond to the KK states and 2-branes of $X$
similarly to \GRT.

% --------------------------------------------------------------------- %
% Maximal decompactifcation limits of $MX(k)$
% --------------------------------------------------------------------- %
\subsec{Maximal decompactifcation limits of $MX(k)$}

Assuming that $X$ exists and that a Hilbert space
description of $X$ compactified on a light-like direction exists
requires the existence of a theory $MX(k)$.
We will now study it in a similar fashion as we studied $X$.

Let us restrict for simplicity to $MX(1)$.
This theory has 8 supersymmetries and our first task is to
find the formula for its BPS excitations.
Since $MX(1)$ describes $X$ compactified on $\MT{5}\times\MR{1,1}$
its Hamiltonian depends on 25 external parameters which describe
the shape size and $B$-fields of the $\MT{5}$.
Thus $MX(1)$ depends on external parameters 
$$
q\in SO(5,5,\BZ) \backslash SO(5,5,\BR) / (SO(5)\times SO(5)).
$$
The BPS particles of $MX(1)$ correspond to longitudinal objects
of $X$. These are either membranes wrapped on one cycle of
$\MT{5}$ or 5-branes wrapped on 4 cycles. They are in the 
$\rep{10}$ of $SO(5,5,\BR)$.
The analysis of the BPS mass formula is identical to the case
of M-theory on $\MT{5}$ analyzed above.
Thus $MX(1)$ has to have particles of masses 
$$
\sqrt{(n_1 r_1)^2 + \sum_2^5 (n_i r_i^{-1})^2},\,
\sqrt{(n_1 r_1)^2 + (n_2 r_2)^2 + \sum_3^5 (n_i r_i^{-1})^2},\,\cdots,\,
\sqrt{\sum_1^5 (n_i r_i)^2}.
$$
We see that $MX(1)$ has two kinds of 5+1D low-energy limits.
Both have $\SUSY{(1,0)}$ supersymmetry. One limit has a low-energy
of one vector multiplet and the other has a low-energy of one
tensor multiplet with moduli space $\MS{1}$.
Both limits have BPS strings with fixed tension.

% --------------------------------------------------------------------- %
% Low-energy description of BPS objects
% --------------------------------------------------------------------- %
\subsec{Low-energy description of BPS objects}

We will proceed to analyze the low-energy description
of strings in $MX(1)$.
We assume that the string is infinite in the 5th direction.
Let the transverse rotation group in directions $1\dots 4$
be denoted by $SO(4)_T$. We will write representations of $SO(4)_T$
as representations of $SU(2)\times SU(2)$.

Since T-duality ($SO(5,5,\BZ)$) along the 5th direction
(when compactified) maps the string to a KK state, the $SO(4)_T$
quantum numbers of the ground state of the string must be
the same as those of its T-dual KK state.

For $\SUSY{(1,0)}$ supersymmetry, the supersymmetry charges
transform in the 
$$
(\rep{2},\rep{4})
$$
of $SU(2)_R\times SO(5,1)$, where $SU(2)_R$ is the R-symmetry.
This R-symmetry is related to the $SO(3)$ R-symmetry of $X$
which rotates the 3 scalars.
Note that both $\rep{2}$ and $\rep{4}$ are pseudo-real
and that one can impose a reality condition on $(\rep{2},\rep{4})$.
Under $SU(2)_R\times SO(4)_T$ the charges transform as
$$
(\rep{2},\rep{2},\rep{1}) +
(\rep{2},\rep{1},\rep{2}).
$$
Let the charges be $Q^{i\a},Q^{i\da}$ the reality condition
is
$$
(Q^{i \a})^\dagger = \epsilon_{ij}\epsilon_{\a\b} Q^{j\b},\qquad
(Q^{i \da})^\dagger = \epsilon_{ij}\epsilon_{\da\db} Q^{j\db}.
$$
Putting a KK state in the 5th direction leaves only 
the $Q^{i\da}$ unbroken. The $Q^{i\a}$ generate zero-modes.
The irreducible representation of the Clifford algebra
of $Q^{i\a}$ decomposes under $SU(2)_R\times SO(4)_T$ as
$$
(\rep{2},\rep{1},\rep{1}) + 
(\rep{1},\rep{2},\rep{1}).
$$
The KK state is in a {\it reducible} representation.
There are two cases to distinguish.
When the 5+1D low-energy is a free vector-multiplet, the KK state
is in
\eqn\kkrv{
(\rep{1},\rep{1},\rep{2}) \times \{
(\rep{2},\rep{1},\rep{1}) + 
(\rep{1},\rep{2},\rep{1})\}
= (\rep{2},\rep{1},\rep{2}) + (\rep{1},\rep{2},\rep{2}).
}
These are the 4 states of a vector and 4 states of gluinos.
When the 5+1D low-energy is a free tensor-multiplet the KK state
is in
\eqn\kkrt{
(\rep{1},\rep{2},\rep{1}) \times \{
(\rep{2},\rep{1},\rep{1}) + 
(\rep{1},\rep{2},\rep{1})\}
= (\rep{2},\rep{2},\rep{1}) + (\rep{1},\rep{3},\rep{1})
+ (\rep{1},\rep{1},\rep{1}).
}
These are the states of an anti-self-dual tensor field, a scalar
and 4 fermions.
By T-duality we deduce that the ground states of the extended
strings are in the opposite representations, i.e. in \kkrv\
when the low-energy is a tensor multiplet and in \kkrt\ when
the low-energy is a vector-multiplet.

How can this multiplicity be realized?
We can expect that the string has a low-energy description
with $\SUSY{(0,4)}$ supersymmetry.
For a single string in $MX(1)$, the low-energy will contain
4 scalars in the $(\rep{2},\rep{2})$ of $SO(4)_T$ and
2 complex 
right-moving fermions in the $(\rep{2},\rep{1})$ of $SO(4)_T$.
The ground states of the Clifford algebra of the zero modes will
then be in the representation
$$
2(\rep{1},\rep{1}) + (\rep{2},\rep{1}).
$$
This means that we need some left-moving fermionic zero modes.
For the case of \kkrt\ their zero-mode algebra should have
two states in the $(\rep{2},\rep{1})$ and for the case of \kkrv\
the zero-mode algebra should have two states in the $(\rep{1},\rep{2})$.
It seems that the unique solution to the problem is to have
three real left-moving fermions which in the case of \kkrt\
are in the $(\rep{3},\rep{1})$ of $SO(4)_T$ and in the case of
\kkrv\ they are in the $(\rep{1},\rep{3})$. Denote these by
$\zeta^{\a\b}$ ot $\zeta^{\da\db}$ respectively.
These are real fermions, symmetric in the spinor indices.
The Clifford algebra is then
$$
\zeta^{\a\b}_0\ket{\gamma} = \epsilon^{\a\gamma}\ket{\b} + 
\epsilon^{\b\gamma}\ket{\a},
$$
or 
$$
\zeta^{\da\db}_0\ket{\dot{\gamma}} = \epsilon^{\da\dot{\gamma}}\ket{\db} + 
\epsilon^{\db\dot{\gamma}}\ket{\da}.
$$

Altogether the low-energy Lagrangian will be for \kkrt:
\eqn\lelagt{
L = \int d^2\sigma \{
\px{+}\phi^a\px{-}\phi^a 
+ \psi^{i\a}\px{+}\psi_{i\a} 
+ \zeta^{\a\b} \px{-}\zeta_{\a\b}\},
}
and for \kkrv:
\eqn\lelagv{
L = \int d^2\sigma \{
\px{+}\phi^a\px{-}\phi^a 
+ \psi^{i\a}\px{+}\psi_{i\a} 
+ \zeta^{\da\db} \px{-}\zeta_{\da\db}\},
}
where $a=1\dots 4$ is a vector index for $SO(4)_T$, $\a=1,2$ and 
$\da =\dot{1},\dot{2}$ are spinor indices
and $i=1,2$ is an R-symmetry index.

The extra left-moving fermions will have an effect on the counting
of $\half$-BPS states in $MX(1)$.
A $\half$-BPS state can be constructed by starting with a string
wound around direction $1$ and exciting some left-moving oscillators
\refs{\SenHBP,\VafGAS,\BSVtop,\VafIOD}.
These are the 4 left-moving scalars and the 3 left-moving fermions.
Thus we expect the multiplicity and $SO(4)_T$ quantum numbers of 
$\half$-BPS states with winding number $w=1$ along direction $1$ and
momentum $p>0$ along direction $1$ to
be according to the states at level $p$ of the
Fock space of states of 4 bosons in the
$(\rep{2},\rep{2})$ of $SO(4)_T$ and 3 fermions in the
$(\rep{3},\rep{1})$ of $SO(4)_T$.
On the other hand, these $\half$-BPS states of $MX(1)$ describe
bound states of one longitudinal 5-brane of $X$ which is wrapped
on the DLCQ direction and on directions $2\dots 5$  together with
$p$ longitudinal 2-branes wrapped on the DLCQ direction and on direction
$1$.
It might be interesting to attempt to quantize that system in the DLCQ
and check this statement.

When there are $k$ extended strings the low-energy description
should be an interacting generalization of \lelagt,\lelagv.
This should be a 1+1D conformal theory with $\SUSY{(0,4)}$
supersymmetry which in a certain limit looks like $k$ copies
of \lelagt\ or \lelagv\ with an $S_k$ orbifold that acts on
all the fields $\phi,\psi,\zeta$.
Although the target space of a 1+1D theory is not an invariant notion
in general, the statement could be made more rigorous if we recast
it in terms of wave-packets.
The interacting theory that we are looking for does not seem to be
easily defined as a perturbation of the orbifold theory.
Unlike the cases of \refs{\DVV,\DVVF},
The twist operator of this orbifold is of weight $(1,{7\over 8})$
and is not only relevant but even not 1+1D Lorenz invariant.

Nevertheless, we will assume that there is an interacting generalization
of \lelagt,\lelagv.

It is perhaps interesting to note that we could have asked a similar
question about the strings of the theory $S(1)$ (of \SeiVBR).
The low-energy description of $k$ strings is a $\SUSY{(4,4)}$
1+1D theory which is very likely to be the theory discussed in \WitQHB.
This was the strongly coupled $\sigma$-model obtained from 
the orbifold $(\MR{4})^k/S_k$ by turning the $\theta$-angle at
the $\BZ_2$ fixed-points to $\theta=0$ rather than $\theta=\pi$.
This is defined by resolving the $\BZ_2$ fixed point locus with
an exceptional divisor and setting the K\"ahler class of that divisor
to zero and the $B$-field to zero as well.

% --------------------------------------------------------------------- %
% Reparameterization anomalies
% --------------------------------------------------------------------- %
\subsec{Reparameterization anomalies}

The 1+1D Lagrangians \lelagt\ and \lelagv\ have
reparameterization anomalies.

The index for the gravitational anomaly for a chiral spinor in the 
representation $\rep{r}$ of a vector-bundle is given by:
$$
\hat{I}_{\rep{r}}(F,R) = \trr{\rep{r}}{e^{i F}} (1 + {1\over {48}}\trp{R^2}
  + \cdots),
$$
where $R$ is the world-sheet curvature and $F$ is the field-strength
of the vector-bundle.
In our case the spinors transform as sections of the normal
bundle and $F$ satisfies:
$$
\trr{\rep{(2,2)}}{F\wdg F} = -\trp{R\wdg R}
$$
For the spinors $\psi$ the representation $\rep{r}$ is $(\rep{2},\rep{1})$
and we find
$$
\hat{I}_{\rep{(2,1)}}(F,R) = {1\over {24}}\trp{R^2}
        -{1\over 8}\trr{\rep{(2,2)}}{F^2} = {1\over 6}\trp{R^2}.
$$
For the spinors $\zeta$ we find
$$
\hat{I}_{\rep{(3,1)}}(F,R) = {1\over {16}}\trp{R^2}
        -{1\over 2}\trr{\rep{(2,2)}}{F^2} = {9\over {16}}\trp{R^2}.
$$
Altogether the anomaly is
$$
({9\over {16}} - {2\over 6}) \trp{R^2} = {{11}\over {48}}\trp{R^2}.
$$
Since the string is not charged under any of the bulk fields,
we cannot cancel the anomaly by an inflow mechanism.

I do not know how to cancel this anomaly.
In flat space one can add $\int_\Sigma \omega_3$
where $\omega_3$ is the 3-form such that $d\omega_3 = \trp{R^2}$ and
$\Sigma$ is any 3-manifold whose boundary is the string.
This extra term will be a $c$-number but will make sure that
the action is independent of the particular coordinate system in which
we chose to write the action down. However, for a curved
5+1D space, it is impossible to add a canonical term like that,
since $\int_\Sigma \omega_3$ depends on $\Sigma$. Maybe this means
that we cannot put $MX$ on a curved background or,
perhaps
this is an indication that one has to add more chiral matter
to the string. 
In fact, without adding more chiral matter \lelagt,\lelagv\ have
a non-vanishing Casimir energy. Since the Casimir energy can be
calculated using only the low-energy Lagrangian its correction
to the mass of a long wound string can be trusted and that
would lead to a contradiction with T-duality.
Nevertheless, in what follows we will assume that $MX$ is well defined.

% --------------------------------------------------------------------- %
% The limit of M(atrix)-theory on $\MT{5}$
% --------------------------------------------------------------------- %
\subsec{The limit of M(atrix)-theory on $\MT{5}$}

When $X$ is compactified on a very small $\MS{1}$, the resulting
5+1D theory is $S(1)$ -- the theory discovered in \SeiVBR.
A M(atrix) model for $S(1)$ has been given in \refs{\ABKSS,\WitQHB}.
It is a 1+1D theory with $\SUSY{(4,4)}$. The $p_\Vert = 1$
sector is given by a free $\SUSY{(4,4)}$ $\sigma$-model with
target space $\MT{4}$.
The $p_\Vert = k$ is given by an interacting
 $\SUSY{(4,4)}$ $\sigma$-model with
target space $(\MT{4})^k/ S_k$ with the singularities resolved
by putting the $\theta$-angle to zero \WitQHB.
The limit of small $\MS{1}$ for $X$ corresponds to the limit of
a large $\MS{1}$ for $MX(k)$. Thus, $MX(k)$ compactified on
$\MT{4}\times\MR{1,1}$ must reduce in the low-energy limit
to the $\SUSY{(4,4)}$ $\sigma$-model found in \refs{\ABKSS,\WitQHB}.
This seems to be the case for $MX(1)$.

% --------------------------------------------------------------------- %
% Properties OF A M(atrix)-model for $MX(k)$.
% --------------------------------------------------------------------- %
\subsec{Properties Of A M(atrix)-model for $MX(k)$}

If $MX(k)$ exists we can compactify one light-like direction and
look for a Hamiltonian which describes the $p_\Vert = N$ sector.
We will denote this theory by $M^2 X(k,N)$.

This has to be a theory with 4 supersymmetries.
The large $N$ limit of it will be a M(atrix)-model for 
$MX(k)$ on $\MT{4}\times\MR{1,1}$ (in one of its low-energy limits).
The theory has to depend on 16 external parameters
$$
q\in SO(4,4,\BZ) \backslash SO(4,4,\BR) / (SO(4)\times SO(4)).
$$
For simplicity, we will proceed with $MX(1)$.
Since the longitudinal strings of $MX(1)$ are unwrapped they are
singlets of $SO(4,4,\BR)$ and their mass is independent of $q$.
Since they are related by the $SO(5,5,\BZ)$ T-duality of
$MX(k)$ to KK states of $MX(k)$, they must have the multiplicity
of KK states. Thus the BPS mass formula for KK states in the
M(atrix)-model for $MX(1)$ (which we denote $M^2 X(1,1)$) is
$$
n,\qquad n\in\BZ^{+}.
$$
This means that $M^2 X(1,1)$ has a 1+1D limit.
This limit must be a conformal theory with 4 supersymmetries.
Since the quantization of $M^2 X(1,1)$ must describe the KK
states of $MX(1)$ we assume that
$M^2 X(1,1)$ has a target space $\MT{4}$.

We believe that the theory $M^2 X(1,1)$ is the $\SUSY{(0,4)}$ theory
given by \lelagv\ (or \lelagt\ for the other low-energy limit of $MX$)
and $M^2 X(k,1)$ is the unknown theory that describes $k$ strings
of $MX(1)$ and was discussed in section (6.2).

Recently, 1+1D theories with $\SUSY{(0,4)}$ have been implemented
\refs{\Lowe,\KSAB}
for the description of the $\SUSY{(1,0)}$ 5+1D theory of \SeiVBR.
These theories are different from the 5+1D theories $MX(k)$
which do not contain hypermultiplets.
Indeed the left-moving fermions of \refs{\Lowe,\KSAB} are
not in the strange $(\rep{3},\rep{1})$ representation as
$\zeta^{\a\b}$ of \lelagt. It might be that the 
interacting theories $M^2 X$ could be defined by modifying
the actions of \refs{\Lowe,\KSAB}.

% --------------------------------------------------------------------- %
% Comment on another approach
% --------------------------------------------------------------------- %
\subsec{A comment on a different approach}

One can try another approach to derive $MX(k)$.
Let us compactify $X$ on $\MT{5}\times\MS{1}$ such that the $\MS{1}$
is in a light-like direction. We need to find a description for
$N$ KK states along $\MS{1}$.
If $\MS{1}$ were space-like
we could have used the $E_{6(6)}(\BZ)$ U-duality to map the KK states
to $N$ 5-branes wrapped on $\MT{5}$.
However, 
we have seen in section (5.5) that the theory of $N$ 5-branes
does not decouple from the bulk $U(1)$. It would be interesting 
to know if one can derive $MX(k)$ from such a setup.

% --------------------------------------------------------------------- %
% Summary of the conjectures
% --------------------------------------------------------------------- %
\subsec{Summary of the conjectures}

Let us summarize the M(atrix) conjectures:

% - - - - - - - - - - - - - - - - - - - - - - - - - - - - - - - - - - - %
% Conjecture 1
% - - - - - - - - - - - - - - - - - - - - - - - - - - - - - - - - - - - %
\bigbreak\bigskip
{\bf $Conjecture\ 1:$}\nobreak

For every $N$ there exists a 5+1D theory $MX(N)$ with 8 supersymmetries
and a low-energy of $N$ tensor multiplets with moduli space
$(\MS{1})^N/ S_N$.
When compactified on $\MT{3}$ the theory has a moduli space of 
 $(\MT{4})^N/S_N$.
When compactified on $\MT{5}$ the theory has an $SO(5,5,\BZ)$ T-duality
group.
The large $N$ limit of this theory is the M(atrix) description of $X$
compactified on $\MT{5}\times\MR{1,1}$.
However, I could not resolve the problem of anomalies and of
the zero-point energy of strings in $MX$.

% - - - - - - - - - - - - - - - - - - - - - - - - - - - - - - - - - - - %
% Conjecture 2
% - - - - - - - - - - - - - - - - - - - - - - - - - - - - - - - - - - - %
\bigbreak\bigskip
{\bf $Conjecture\ 2:$}\nobreak

For every $k,N$ there exists a 1+1D theory 
$M^2 X(k,N)$ with $\SUSY{(0,4)}$
supersymmetry and an $SO(4,4,\BZ)$ T-duality
group.
The large $k$ limit of this theory is the M(atrix) description of $MX(k)$
compactified on $\MT{4}\times\MR{1,1}$.
%%% Further, we conjecture that $MX(1,1)$ is the free theory given
%%% by \lelagt.

% ===================================================================== %
% Section (): Discussion
% ===================================================================== %
\newsec{Discussion}

We have argued that if there exists a M(atrix) model which describes
the DLCQ sector with $p_\Vert = 1$ of M-theory on $\MT{6}$ it must
have a 6+1D decompactified limit.
This will be a local 6+1D theory without gravity,
whose low-energy is $U(1)$ SYM.
In units for which the $U(1)$ coupling constant is of order one,
this theory has 2-branes and 5-branes with finite tension.
Existence of such a local theory implies the existence of 
5+1D conformal theories with $\SUSY{(1,0)}$ supersymmetry and
$k$ tensor multiplets with a moduli space of $\MR{k}/S_k$.

We have briefly discussed the possible cancelation of
reparameterization anomalies for the chiral theories which
arise in the low-energy description of curved branes.
We suggested that the 5-brane anomaly might be canceled by
a 6+1D bulk term $\int J_7$ -- a mechanism which, for
curved 6+1D space-time, is only possible
when the world-volume is of co-dimension one, as in our case.

We have explored the possibility that $X$ could be described by a 
M(atrix) model of its own -- denoted $MX$.
We argued that such a theory should be a 5+1D theory with 
8 supersymmetries and have a $SO(5,5,\BZ)$ T-duality.
Its M(atrix) model should then be a 1+1D theory with $\SUSY{(0,4)}$
supersymmetry. We presented a conjectured free theory for the
$p_\Vert = 1$ sector of the DLCQ.

We showed that T-duality implies that there are macroscopic string
states in $MX$, but we had a problem with their
reparameterization anomaly and the zero-point energy.
I do not know how it is canceled.

If a theory such as $X$ really exists, it might be useful for studying
field-theories since it has higher dimensional branes with only 
8 supersymmetries.

Can $X$ and $MX$ be realized in M-theory?
First we point out that since M-theory is not fully understood
we do not know the full extent of the notion ``realized in M-theory''.
Up until recently, the only method to extract field theories out of
M-theory was via low-energy questions. In \SeiVBR\ a new method for
extracting a subset of degrees of freedom has been presented.
Perhaps the anomaly cancelation mechanism discussed in (5.6)
might provide a clue on how to embed $X$ in M-theory or
could be a starting point for a No-Go theorem.
In any case, if $X$ exists it is realized in at least one way in
M-theory -- by compactifying a light-like direction!
It would seem that the M(atrix) principle, namely that theories
should have a DLCQ Hamiltonian is more important than the hope
that all theories should be realized in a geometrical setting of
what is now known about M-theory.

% ===================================================================== %
% Acknowledgments
% ===================================================================== %
\bigbreak\bigskip\bigskip
\centerline{\bf Acknowledgments}\nobreak
I am indebted to D. Kutasov, L. Susskind and E. Witten and in particular
to N. Seiberg for very helpful discussions and comments.
I am especially grateful to S. Sethi for collaborating in early
stages of this work and for many useful discussions.

This research was supported by a Robert H. Dicke fellowship and by
DOE grant DE-FG02-91ER40671.

\listrefs
\bye
\end